\pdfoutput=1
\documentclass[letterpaper, 10 pt, conference]{ieeeconf}  

\IEEEoverridecommandlockouts                              

\usepackage{graphicx}
\usepackage{url}
\usepackage{times}
\usepackage{epstopdf}
\usepackage{algorithmic}
\usepackage{amsmath} 
\usepackage{amssymb}  
\usepackage{balance}

\newcommand{\E}{{\bf E}}

\newtheorem{theorem}{Theorem}

\newtheorem{definition}[theorem]{Definition}

\newcommand{\eat}[1]{}

\makeatletter
\def\@begintheorem#1#2{\sl \trivlist \item[\hskip \labelsep{\bf #1\ #2:}]}
\def\@opargbegintheorem#1#2#3{\sl \trivlist
      \item[\hskip \labelsep{\bf #1\ #2\ #3:}]}
\makeatother

\title{\LARGE \bf
A Model for Learned Bloom Filters and Related Structures
}

\author{Michael Mitzenmacher{$^1$}\thanks{{$^1$}School~of Engineering and Applied Sciences, Harvard University.  
Supported in part by NSF grants CCF-1563710, CCF-1535795, and CCF-1320231.
This work was done while visiting Microsoft Research.}}

\begin{document}

\maketitle
\thispagestyle{empty}
\pagestyle{empty}

\begin{abstract}
Recent work has suggested enhancing Bloom filters by using a
pre-filter, based on applying machine learning to model the data set
the Bloom filter is meant to represent.  Here we model 
such {\em learned Bloom filters}, clarifying what guarantees can and cannot be
associated with such a structure. 
\end{abstract}

\section{Introduction}

An interesting paper, ``The Case for Learned Index
Structures'' \cite{TCFLIS}, recently appeared, suggesting that
standard index structures and related structures, such as Bloom
filters, could be improved by using machine learning to develop what
they authors dub learned index structures.  Here we aim to provide a
more formal model for their variation of a Bloom filter, which they
call a {\em learned Bloom filter}, and clarify what features it does
and does not have.  A key issue is that, unlike standard Bloom
filters, the performance of learned Bloom filters depends on both the data
set the Bloom filter represents and the set membership queries made.
Because of this, the types of guarantees offered by learned Bloom
filters differ significantly from those of standard Bloom filters.  We
formalize this issue below.

The performance of learned Bloom filters will therefore necessarily be
application dependent.  Indeed, there may be applications where they
offer significant advantages over standard Bloom filters.  We view
this work as a beginning step in laying out a theoretical framework to
understand learned Bloom filters and related structures, with a
particular goal of pointing out issues that may affect their
suitability for real-world applications.  

In what follows, we start by reviewing standard Bloom filters and variants,
following the framework provided by the reference \cite{BroderMitzenmacher}.
We then describe learned Bloom filters, and attempt to provide a model 
which highlights both their potential benefits and their limitations.
In particular, we find learned Bloom filters appear most useful when the
query stream can be modelled as coming from a fixed distribution,
which can be sampled during the construction.  

\section{Review:  Bloom Filters}

\subsection{Definition of the Data Structure}

A Bloom filter for representing a set $S = \{x_1,x_2,\ldots,x_n\}$ of
$n$ elements corresponds to an array of $m$ bits, and uses $k$
independent hash functions $h_1,\ldots,h_k$ with range
$\{0,\ldots,m-1\}$.  Typically we assume that these hash functions are
perfect; that is, each hash function maps each item in the universe to
independently and uniformly to a number in $\{0,\ldots,m-1\}$.
Initially all array bits are 0.  For each element $x \in S$, the array
bits $h_i(x)$ are set to 1 for $1 \leq i \leq k$; it does not matter
if some bit is set to 1 multiple times.  To check if an item $y$ is in
$S$, we check whether all $h_i(y)$ are set to 1.  If not, then clearly
$y$ is not a member of $S$.  If all $h_i(y)$ are set to 1, we conclude
that $y$ is in $S$, although this may be a {\em false positive}.  A
Bloom filter does not produce false negatives.

There are various theoretical guarantees one can associate with a
Bloom filter.  The simplest is the following.  Let $y$ be an element
of the universe such that $y \notin S$, where $y$ is chosen
independently of the hash functions used to create the filter.  A
useful way to think of this is that an adversary can choose any
element $y$ before the Bloom filter is constructed; the adversary has
no knowledge of the hash functions used, but may know the set $S$.  Let $\rho$ be the fraction
of bits set to 1 after the elements are hashed.  Then
$$\Pr(y \mbox{ yields a false positive}) = \rho^k.$$ 
Further, probabilistic analysis shows both that
$$\E[\rho] = 1 - \left ( 1 - {1 \over m} \right )^{kn} \approx 1-{\rm e}^{-kn/m},$$
and that 
$$\Pr(|\rho - \E[\rho]| \geq \gamma) \leq {\rm e}^{-\Theta(\gamma^2 m)}$$ in
the typical regime where $m/n$ and $k$ are constant.  That is, $\rho$
is, with high probability, very close to its easily calculable
expectation, and thus we know (up to very small random deviations, and
with high probability over instantiations of the Bloom filter) what
the probability is than an element $y$ will be a false positive.
Because of this, it is usual to talk about the {\em false positive
probability} of a Bloom filter; in particular, it is generally
referred to as though it is a constant depending on the filter
parameters, although it is a random variable, because it is tightly
concentrated around its expectation.  

Moreover, given a set of distinct query elements $Q
= \{y_1,y_2,\ldots,y_q\}$ with $Q \cap S = \emptyset$ chosen a priori
before the Bloom filter is instantiated, the fraction of false
positives over these queries will similarly be concentrated near
$\rho^k$.  Hence we may talk about the {\em false positive rate}
of a Bloom filter, which (when the query elements are distinct) is essentially the
same as the false positive probability.  (When the query elements are
not distinct, the false positive rate may vary significantly,
depending on on the distribution of the number of appearances of
elements and which ones yield false positives; we focus on the distinct
item setting here.)  In particular, the false positive rate is a
priori the same for {\em any} possible query set $Q$.  Hence one
approach to finding the false positive rate of a Bloom filter
empirically is simply to test a random set of query elements (that
does not intersect $S$) and find the fraction of false positives.
Indeed, it does not matter what set $Q$ is chosen, as long as it is
chosen independently of the hash functions.

We emphasize that, as we discuss further below, the term false positive
rate often has a different meaning in the context of learning theory 
applications.  This difference of terminology is a possible point of
confusion in \cite{TCFLIS}, and care must be taken in understanding how
the term is being used.  

\subsection{Additional Bloom Filter Benefits and Limitations}

For completeness, we relate some of the other benefits and limitations of
Bloom filters.  More details can be found in \cite{BroderMitzenmacher}.

We have assumed in the above analysis that the hash functions are
fully random.  As fully random hash functions are not practically
implementable, there are often questions relating to how well the
idealization above matches the real world for specific hash functions.
In practice, however, the model of fully random hash functions appears
reasonable in many cases; see \cite{Simple} for further discussion on
this point.

If an adversary has access to the hash functions used, or to the final
Bloom filter, it can find elements that lead to false positives.  One
must therefore find other structures for adversarial situations.  A
theoretical framework for such settings is developed in \cite{naor2015bloom}.
Variations of Bloom filters, which adapt to false positives
and prevent them in the future, are described in \cite{adapt2,adapt1};  while not meant
for adversarial situations, they prevent repeated false positives with the
same element.  

One of the key advantages of a standard Bloom filter is that it is
easy to insert an element (possibly slightly changing the false
positive probability), although one cannot delete an element without
using a more complex structure, such as a counting Bloom filter.  
However, there are more recent alternatives to the standard Bloom filter,
such as the cuckoo filter \cite{cuckoof}, which can achieve the same or better space 
performance as a standard Bloom filter while allowing insertions and deletions.
If the Bloom filter does not need to insert or delete elements, a well-known
alternative is to develop a perfect hash function for the data set, and store
a fingerprint of each element in each corresponding hash location (see, e.g., \cite{BroderMitzenmacher}
for further explanation);  this approach reduces the space required by approximately 30\%.  

\section{Learned Bloom Filters}

\subsection{Definition of the Data Structure}

We now consider the learned Bloom filter construction offered in
\cite{TCFLIS}.  We are given a set of positive keys ${\cal K}$ that correspond
to set to be held in the Bloom filter -- that is, ${\cal K}$ corresponds
to the set $S$ in the previous section.  We are also given a set of
negative keys ${\cal U}$ for training.  We then train a neural network with
${\cal D} = \{(x_i,y_i =1)~|~x_i \in {\cal K}\} \cup
\{(x_i,y_i =0)~|~x_i \in {\cal U}\}$;  that is, they suggest
using a neural network on this
binary classification task to produce a probability, based on minimizing
the log loss function
$$L = \sum_{(x,y) \in {\cal D}} y \log f(x) + (1-y)\log (1-f(x)),$$
where $f$ is the learned model from the neural network.  Then $f(x)$
can be interpreted as a probability that $x$ is a key from the set.
Their suggested approach is to choose a threshold $\tau$ so that if
$f(x) \geq \tau$ then the algorithm returns that $x$ is in the set,
and no otherwise.  Since such a process may provide false negatives
for some keys in ${\cal K}$ that have $f(x) < \tau$, a secondary
structure -- such as a smaller standard Bloom filter for such keys --
can be used to ensure there are no false negatives, thereby matching
this important characteristic of the standard Bloom filter.

In essence, \cite{TCFLIS} suggests using a pre-filter ahead of the
Bloom filter, where the pre-filter comes from a neural network and
estimates the probability an element is in the set, allowing the use
of a smaller Bloom filter.  Performance improves if the size to
represent the learned function $f$ and the size of the smaller backup
filter for false negatives is smaller than the size of a corresponding
Bloom filter with the same false positive rate.  While the idea of
layering multiple filters has appeared in previous work, this approach
appears novel.  Indeed, the more typical setting is for the Bloom
filter itself to be used as a pre-filter for some other, more
expensive filtering process.  Of course the pre-filter here need not
come from a neural network; any approach that would estimate the
probability an input element is in the set could be used.

\subsection{Defining the False Positive Probability:  High Level Issues}

The question remains how to determine or derive the false positive
probability for such a structure, and how to choose an appropriate
threshold $\tau$.  One approach would be to empirically 
find the false positive rate over a test set, and this appears to be
what has been done in \cite{TCFLIS}.  This approach is, as we
have stated, suitable for a standard Bloom filter, where the false
positive rate is guaranteed to be close to its expected value for any
test set, with high probability.  But as we explain in the next subsection, this
methodology requires significant additional assumptions in the learned Bloom 
filter setting.

Before formalizing appropriate definitions, to frame the issue it is
helpful to think of an adversarial situation, although as we discuss
below an adversary is not strictly necessary.  An adversary might
naturally be able to find items for which $\Pr(f(y) \geq \tau)$ is
surprisingly large based on their own analysis of the data, even
without access to the structure $f$ that is finally determined.  That
is, consider the following intuition. The function $f$ is designed to
to take advantage of structure in the set ${\cal K}$, as well as the
information in the collection of non-set elements ${\cal U}$.  An
adversary, knowing ${\cal K}$ and/or ${\cal U}$ themselves, might be
able to design elements that are similar to the elements of ${\cal
K}$.  An element similar to the elements of ${\cal K}$ should have a
large $f(y)$ value, and hence be more likely to yield a false
positive.

More formally, let us assume the adversary knows ${\cal K}$ and/or
${\cal U}$, and chooses an element $y \notin {\cal K}$ to test.
In the standard Bloom filter setting, the array of bits that constitutes
the filter is generated from random hash values, and so knowing the
set $S$ tells the adversary nothing about what $y$ value might be most
likely to yield a false positive.  All $y$ values are equivalent (as
long as $y \notin S$).  In the learned Bloom filter setting, knowing
${\cal K}$ and ${\cal U}$ may give the adversary information about the
resulting index function $f$, even if this knowledge is just that $f$
is designed to give higher values to elements similar to ${\cal K}$
and lower values to elements in ${\cal U}$.  The adversary need not
even know the specific method used to determine $f$; the knowledge of
the data alone may allow the adversary to choose a $y$ value with a
large expected $f(y)$ value, so that $\Pr(f(y) \geq \tau) > \epsilon$;
that is the probability of a false positive is larger than expected.

While we have stated this problem in terms of an adversary, one does
not need to posit an adversary to see that this situation might arise
naturally in standard data scenarios.  All we need to suppose is that
the query set ${\cal Q}$ of elements that one uses to query the Bloom filter
are similar to the set ${\cal K}$ in some manner that might be
captured by $f$, so that the expected value of $f(y)$ for $y \in {\cal Q}$
skews large.  (Here we assume ${\cal Q}$ is disjoint from ${\cal K}$, as
we are interested in the rate of false positives, not true positives.)

An example based on ranges may be helpful.  Suppose the universe of
elements is the range $[0,1000000)$, and the set ${\cal K}$ of elements to
store in our Bloom filter consists of a random subset of elements from
the range $[1000,2000]$, say half of them, and 500 other random elements
from outside this range.  Our learning algorithm might determine that a
suitable function $f$ is $f(y)$ is large (say $f(y) \approx 1/2$) for
elements in the range $[1000,2000]$ and close to zero elsewhere, and
then a suitable threshold might be $\tau = 0.4$.  The resulting false
positive rate will depend substantially on what elements are queried.
If ${\cal Q}$ consists of elements primarily in the range $[1000,2000]$, the
false positive rate will be quite high, while if ${\cal Q}$ is chosen 
uniformly at random over the whole range, the false positive rate will
be quite low.  The main point is that the false positive rate, unlike
in the setting of a standard Bloom filter, is highly dependent on the
query set, and as such is not well-defined independently of the queries,
as it is for a standard Bloom filter.

Indeed, it seems plausible that in many situations, the query set ${\cal Q}$
might indeed be similar to the set of elements ${\cal K}$, so that
$f(y)$ for $y \in {\cal Q}$ might often be above naturally chosen thresholds.
For example, in security settings, one might expect that queries for
objects under consideration (URLs, network flow features) would be
similar to the set of elements stored in the filter.  The key here is
that, unlike in the Bloom filter setting, the false positive
probability for a query $y$ can depend on $y$, even before the ``data
structure'', which in this case corresponds to the function $f$, is
instantiated.

It is worth noting, however, that the problem we point out here can
possibly be a positive feature in other settings; it might be that the false
positive rate is remarkably low if the query set is suitable.
Again, one can consider the range example above where queries are
uniform over the entire space;  the query set is very unlikely to hit
the range where the learned function $f$ yields an above threshold
value in that setting for an element outside of ${\cal K}$.  More generally, one may have query
sets ${\cal Q}$ where the values $f(y)$ for $y \in {\cal Q}$ are smaller than one
might expect.  The key point again remains that the false positive
probability is dependent on the data and the query in what may not 
be predictable ways, in sharp contrast to standard Bloom filters.

\subsection{Defining the False Positive Probability, and Analysis from Empirical Data}

We can formalize (at least partially) settings where we can obtain 
good performance from a learned Bloom filter, given enough data.
The framework below follows standard lines, but provides definitions
to capture the high level ideas given above.
We first formalize the construction of \cite{TCFLIS}.  

\begin{definition}
A {\em learned Bloom filter} on a set of positive keys ${\cal K}$ and
negative keys ${\cal U}$ is a function $f:U \rightarrow [0,1]$ and
threshold $\tau$, where $U$ is the universe possible query keys, and
an associated standard Bloom filter $B$, referred to as a backup filter.  The 
backup filter is set to hold the set of keys $\{z: z \in {\cal K}, f(z) < \tau\}$.  
For a query $y$, the learned Bloom filter returns that $y \in {\cal K}$
if $f(y) \geq \tau$, or if $f(y) < \tau$ and the backup filter
returns that $y \in {\cal K}$.  The learned Bloom filter returns 
$y \notin {\cal K}$ otherwise.
\end{definition}

Note that the size of a learned Bloom filter corresponds to size used to represent
the function $f$ and the size of the backup filter $B$, which we denote by $|f|+|B|$.
In cases where the learned Bloom filter is effective, one expects $f$ to have a small
representation, and the number of false negatives from ${\cal K}$ in the backup filter
to be a reasonably small fraction of ${\cal K}$.

The learned Bloom filter as defined has no false negatives, due to the
backup filter.  We can define the false positive rate of a learned Bloom 
filter with respect to a given query distribution.  

\begin{definition}
A {\em false positive rate on a query distribution} ${\cal D}$ 
over $U - {\cal K}$ for a learned Bloom filter $(f,\tau,B)$ is given by 
$$\Pr_{y \sim {\cal D}}(f(y) \geq \tau)  + (1-\Pr_{y \sim {\cal D}}(f(y) \geq \tau))F(B),$$
where $F(B)$ is the false positive rate of the backup filter $B$.
\end{definition}
While technically $F(B)$ is itself a random variable, as discussed previously, the false
positive rate is well concentrated around its expectations, which depends only on the size
of the filter $|B|$ and the number of false negatives from ${\cal K}$ that must be stored in
the filter, which depends on $f$.  Hence we may naturally refer to the false positive rate
of the learned Bloom filter as being determined by $f$, $\tau$, and $|B|$ rather than on $B$
itself.  That is, where the meaning is clear, we may consider 
the false positive rate on a query distribution for a learned Bloom filter with associated
$(f,\tau)$ to be 
$$\Pr_{y \sim {\cal D}}(f(y) \geq \tau)  + (1-\Pr_{y \sim {\cal D}}(f(y) \geq \tau))\E[F(B)],$$
where the expectation $\E[F(B)]$ is meant to over instantiations of the Bloom filter with given size $|B|$.

Given sufficient data, we can determine an {\em empirical false
positive rate} on a test set, and use that to predict future behavior.
Under the assumption that the test set has the same distribution as
future queries, standard Chernoff bounds provide that the empirical
false positive rate will be close to the false positive rate on future
queries, as both will be concentrated around the expectation.  In many
learning theory settings, this empirical false positive rate appears to
be referred to as simply the false positive rate;  we emphasize that false
positive rate, as we have explained above, typically means something different
in the Bloom filter literature.

\begin{definition}
The {\em empirical false positive rate on a set} ${\cal T}$, where
${\cal T} \cap {\cal K} = \emptyset$, 
for a learned Bloom filter $(f,\tau,B)$ is the number of false positives
from ${\cal T}$ divided by $|{\cal T}|$.
\end{definition}

\begin{theorem}
\label{thmone}
Consider a learned Bloom filter $(f,\tau,B)$, a test set ${\cal T}$,
and a query set ${\cal Q}$, where ${\cal T}$ and ${\cal Q}$ are both determined
from samples according to a distribution ${\cal D}$.  
Let $X$ be the
empirical false positive rate on $|{\cal T}|$, and $Y$ be the
empirical false positive rate on ${\cal Q}$.  Then 
$$\Pr(|X - Y| \geq \epsilon) \leq {\rm e}^{-\Omega(\epsilon^2 \min(|{\cal T}|,|{\cal Q|}))}.$$
\end{theorem}

\smallskip

\begin{proof}
Let $\alpha = \Pr_{y \sim {\cal D}}(f(y) \geq \tau)$, and $\beta$ be 
false positive rate for the backup filter.  We first 
show that for ${\cal T}$ and $X$ that
$$\Pr(|X - (\alpha + (1-\alpha)\beta)| \geq \epsilon) \leq 2{\rm e}^{-\epsilon^2|{\cal T}|}.$$
This follows from a direct Chernoff bound (e.g., \cite{MU}[Exercise 4.13]), since each sample chosen
according to ${\cal D}$ is a false positive with probability $\alpha + (1-\alpha)\beta$.  
A similar bound holds for ${\cal Q}$ and $Y$.

We can therefore conclude 
\begin{eqnarray*}
\Pr(|X - Y| \geq \epsilon) \! \! & \leq & \! \! \Pr(|X - (\alpha + (1-\alpha)\beta)| \geq \epsilon/2) \\
                           \! \! & & \! \! \mbox{ } + \Pr(|Y - (\alpha + (1-\alpha)\beta)| \geq \epsilon/2) \\
                           \! \! & \leq & \! \! 2{\rm e}^{-\epsilon^2|{\cal T}|/4} + 2{\rm e}^{-\epsilon^2|{\cal Q}|/4},
\end{eqnarray*}
giving the desired result.
\end{proof}

Theorem~\ref{thmone} also informs us that it is reasonable to find a suitable parameter $\tau$,
given $f$,  by trying a suitable finite discrete set of values for $\tau$,
and choosing the best size-accuracy tradeoff for the application.
By a union bound, all choices of $\tau$ will perform close to their expectation
with high probability.

We note that while Theorem~\ref{thmone} requires the test set and
query set to come from the same distribution ${\cal D}$, the negative
examples ${\cal U}$ do not have to come from that distribution.  Of
course, if negative examples ${\cal U}$ are drawn from ${\cal D}$, it
may yield a better learning outcome $f$.  

If the test set and query set distribution do not match, because for
example the types of queries change after the original gathering of
test data ${\cal T}$, Theorem~\ref{thmone} offers limited guidance.
Suppose ${\cal T}$ is derived from samples from distribution ${\cal
D}$ and ${\cal Q}$ from another distribution ${\cal D'}$. If the two
distributions are close (say in $L_1$ distance), or, more
specifically, if the changes do not significantly change the
probability that a query $y$ has $f(y) \geq \tau$, then the empirical
false positive rate on the test set may still be useful.  However, in
practice it may be hard to provide such guarantees on the nature of
future queries.  This explains our previous statement that learned
Bloom filters appear most useful when the query stream can be modelled
as coming from a fixed distribution, which can be sampled during the
construction.

We can return to our previous example to understand these effects.
Recall our set of elements is a random subset of half the elements
from the range $[1000,2000]$ and 500 other random elements from the
range $[0,1000000)$.  Our learned Bloom filter has $f(y) \geq \tau$
for all $y$ in $[1000,2000]$ and $f(y) < \tau$.  Our back filter will
therefore store 500 elements.  If our test set is uniform over
$[0,1000000)$ (excluding elements stored in the Bloom filter), our
false positive rate from elements with too large an $f$ value would be
approximately $0.0002$; one could choose a back filter with roughly the
same false positive probability for a total empirical false positive
probability of $0.0004$.  The size of the backup filter would need to
be slightly larger than half the size of a standard Bloom filter
achieving a false positive probability of $0.0004$; although it holds
half the elements, it must achieve half the positive rate, adding
almost 1.5 extra bits per element stored.  If, however, our queries
are uniform over a restricted range $[0,100000)$, then the false positive
probability would jump to $0.0022$ for the learned Bloom filter.

\subsection{Additional Learned Bloom Filter Benefits and Limitations}

Learned Bloom filters can easily handle insertions into ${\cal K}$ by
adding the element, if is does not already yield a (false) positive,
to the backup filter.  Such changes have a larger effect on the false
positive probability than for a standard Bloom filter, since the
backup filter is smaller.  Elements cannot be deleted naturally from a
learned Bloom filter.  A deleted element would simply become a false
positive, which (if needed) could possibly be handled by an additional
structure.

As noted in \cite{TCFLIS}, it may be possible to re-learn a new
function $f$ if the data set changes substantially via insertions and
deletion of elements from ${\cal K}$.  Of course, besides the time needed to
re-learn a new function $f$, this requires storing
the original set somewhere, which may not be necessary for alternative schemes.
Similarly, if the false positive probability proves higher than
desired, one can re-learn a new function $f$; again, doing so will
require access to ${\cal K}$, and maintaining a (larger) set ${\cal U}$
of negative examples.  

\section{Conclusion}

The recent work on learned index structures \cite{TCFLIS}, including the learned
Bloom filter, appears to be generating interest, and is well worthy of
further attention.  However, in order to properly compare the learned
index approach against other approaches, it will prove useful to
develop a suitable theoretical foundation for understanding their
performance, in order to better recognize where the approach can
provide gains and to avoid possible pitfalls.  Here we have attempted
to clarify a particular issue in the Bloom filter setting, namely the
dependence of what is referred to as the false positive rate
in \cite{TCFLIS} on the query set, and how it might affect the
applications this approach is suited for.  This discussion is meant to
encourage users to take care to make sure they realize all of the
implications of the new approach before adopting it.  In particular,
we point out that one should also consider variations on Bloom filters
for comparison;  the cuckoo filter, in particular, already uses less space
than standard Bloom filters (for reasonable false positive rates), has 
similar theoretical guarantees, and allows for insertions and deletions.  

We hope richer theoretical foundations may follow.  
Future work may consider relaxing requirements on the relationship 
between the test set and the query set while achieving some form of
guarantee, or, for specific settings, trying to formally prove the
behavior of the learned function $f$.  Known approaches, based
on for example VC dimension or Rademacher complexity, may apply, 
although the setting is slightly different than many learning applications
in that there is a ``fixed'' set of positive instances that are initially
given in the Bloom filter setting.

\section{Acknowledgments}

The author thanks Suresh Venkatasubramanian for suggesting a closer look
at \cite{TCFLIS}, and thanks the authors of \cite{TCFLIS} for helpful discussions
involving their work.

\end{document}